%% file: arena2012_preprint.tex
\documentclass[a4paper]{article}
\usepackage{graphicx}
\usepackage[numbers]{natbib}
\usepackage{authblk}

\newcommand{\secref}[1]{Section~\ref{#1}}
\newcommand{\figref}[1]{Figure~\ref{#1}}

\newcommand{\url}[1]{\texttt{#1}}

\begin{document}
\input{journals}

\title{LUNASKA neutrino search with the Parkes and ATCA telescopes}
\date{}

\newcommand{\adelaideuni}{\small School of Chemistry \& Physics, Univ. of Adelaide, SA 5005, Australia}
\newcommand{\atnf}{\small CSIRO Astronomy and Space Science, Epping, NSW 1710, Australia}
\newcommand{\erlangenuni}{\small ECAP, Univ.\ of Erlangen-Nuremberg, 91058 Erlangen, Germany}
\newcommand{\astron}{\small ASTRON, 7990 AA Dwingeloo, The Netherlands}

\author[1,2]{J.D. Bray}
\author[2]{R.D. Ekers}
\author[1]{R.J. Protheroe}
\author[3]{C.W. James}
\author[2]{C.J. Phillips}
\author[2]{P. Roberts}
\author[2]{A. Brown}
\author[2]{J.E. Reynolds}
\author[4]{R.A. McFadden}
\author[1]{M. Aartsen}

\affil[1]{\adelaideuni}
\affil[2]{\atnf}
\affil[3]{\erlangenuni}
\affil[4]{\astron}

\maketitle

\begin{abstract}
 The Moon is used as a target volume for ultra-high energy neutrino searches with terrestrial radio telescopes.  The LUNASKA project has conducted observations with the Parkes and ATCA telescopes; and, most recently, with both of them in combination.  We present an analysis of the data obtained from these searches, including validation and calibration results for the Parkes-ATCA experiment, as well as a summary of prospects for future observations.
\end{abstract}

\section{Introduction}

The use of the Moon as a target volume for ultra-high energy (UHE) neutrino searches, monitored with terrestrial radio telescopes, was originally suggested in 1989 by \citet{dagkesamanskii1989}.  The detection principle is to search for the radio pulse, produced via the Askaryan effect~\citep{askaryan1962}, from a neutrino-initiated particle cascade in the lunar regolith.  This technique was first applied with the Parkes radio telescope in 1995 by \citet{hankins1996}, and has been the focus of an increasing tempo of experiments in recent years~\citep{gorham2004a,beresnyak2005,james2010,spencer2010,jaeger2010,buitink2010}.

Compared to experiments searching for radio emission produced via the same mechanism by neutrino-initiated cascades in Antarctic ice~\citep{kravchenko2011,gorham2010}, the maximum aperture for neutrino detection is larger due to the greater detector volume within the field of view of the radio instrument; but the minimum threshold neutrino energy is higher, as an interacting neutrino must be particularly energetic to produce a radio pulse which can be detected at the distance of the Moon.  These properties are further dependent on the radio frequency at which the instrument operates: experiments using low frequencies (100--300~MHz), at which the radio emission from the cascade is weaker but more broadly beamed, have still larger apertures and higher thresholds than experiments using high frequencies (1--3~GHz).  These lunar experiments are therefore best suited to detecting neutrino spectra which extend beyond energies of $\sim 10^{21}$~eV (for high frequencies) or $\sim 10^{22}$~eV (for low frequencies).  Such neutrino spectra are not expected from interactions of cosmic rays (CRs) with the cosmic microwave background~\citep{allard2006}, but may arise from top-down models of UHE CRs, in which they may originate from topological defects such as cusps and kinks in cosmic strings~\citep{berezinsky2011,lunardini2012}.  In the absence of neutrinos from top-down models, lunar experiments are likely to, as they become more sensitive, detect CRs directly before they detect neutrinos~\citep{jeong2012}.

Much work has been done to determine the effective neutrino apertures of lunar experiments, but significant uncertainties still remain.  The analytic calculations of \citet{gayley2009} are reasonably consistent with the Monte Carlo simulations of \citet{beresnyak2003} and \citet{james2009b}, while the simulations of \citet{gorham2001} and \citet{scholten2006} are around an order of magnitude more optimistic.  At high frequencies, the effects of small-scale lunar surface roughness are important, as it scatters the radio emission across a range of directions: \citet{james2010} find that this leads to a small decrease in aperture near the threshold energy, but a large improvement \mbox{($\sim 20\times$)} at higher energies.  The aperture is also strongly dependent on the neutrino-nucleon cross-section, which is poorly constrained at energies above \mbox{$\sim 10^{18}$~eV}, beyond the reach of particle accelerator experiments.  UHE neutrinos may, in addition to the usual weak interactions, interact via the formation and subsequent decay of a microscopic black hole, which leads to a larger cross-section than expected from standard model physics~\citep{anchordoqui2003,connolly2011}.  \Citet{james2010} and \citet{jeong2012} find that this strongly enhances the aperture of lunar neutrino experiments; consequently, some combinations of models of enhanced cross-sections (from black hole formation) and enhanced neutrino fluxes (from top-down UHE CR origin models) may already be excluded by current experiments.

\subsection{LUNASKA experiments}

The sensitivity of a lunar experiment to neutrinos depends on the sensitivity of the radio telescope with which it is performed.  The capabilities of radio telescopes are improving rapidly, which allows greater sensitivity to neutrinos without requiring the development of expensive dedicated instruments.  Much of this effort is directed towards the Square Kilometre Array (SKA)\footnote{\url{http://www.skatelescope.org/}}, an international radio telescope with a total collecting area of 1~km$^2$ scheduled to begin construction in 2016.  The LUNASKA (Lunar Ultra-high energy Neutrino Astrophysics with the SKA) project aims to develop the theory and experimental practice required to perform a lunar neutrino experiment with the SKA when it is complete, through a series of experiments with existing instruments.

The first LUNASKA experiment was conducted in 2008 with three 22~m antennas of the Australia Telescope Compact Array (ATCA), operating in the frequency range 1.2--1.8~GHz.  No coincident radio pulses from the Moon were detected in 33.5 hours of observations, allowing a limit to be placed on the UHE neutrino flux~\citep{james2010}.

The second LUNASKA experiment used the 64~m antenna of the Parkes radio telescope, observing for 173.5 hours in 2010.  This experiment used the Parkes 21~cm multibeam receiver, which has a frequency range of 1.2--1.5~GHz.  An anticoincidence filter between its separate beams was applied, and was successful in excluding the background of radio frequency interference (RFI).

The most recent LUNASKA experiment was conducted in 2011, using both of the above telescopes.  The Parkes radio telescope was used as in the previous experiment; and, in addition, also triggered the storage of buffered data on five antennas of the ATCA.  Due to an upgrade~\citep{mcclure-griffiths2011}, the ATCA frequency range was expanded to 1.1--3.1~GHz, making it sufficiently sensitive to confirm or reject a possible detection with the Parkes radio telescope.

In this article, we provide an update on our experiment with the Parkes radio telescope, describing the signal-processing strategies used to maximize the sensitivity (\secref{sec:parkes_signal}), and the successful rejection of the RFI background (\secref{sec:parkes_results}).  We also present preliminary results from the combined Parkes-ATCA experiment, showing the specialized calibration required for this application (\secref{sec:parkes-atca}), and describe the prospects for lunar neutrino experiments with some future radio instruments (\secref{sec:future}).

\section{Signal optimization for the Parkes experiment}
\label{sec:parkes_signal}

The properties of the radio pulse from a particle cascade are known from simulation and experiment, and the dominant noise in this type of experiment --- primarily thermal radiation from the Moon --- is extremely Gaussian (though RFI is an exception to this: see \secref{sec:parkes_results}).  Under these conditions, the signal-to-noise ratio is optimized by passing the signal through two filters: a pre-whitening filter, which gives the noise a flat spectrum; followed by a matched filter, which matches the expected shape of the pulse (see e.g. \citep{byrne2005}, ch.\ 42).  These two filters can be combined into a single filter, which is fully described by its transfer function in the frequency domain.  The required characteristics of the filter can be conceptually separated as different properties of its transfer function.

The amplitude of the transfer function represents the optimization of the filter bandpass to match the spectrum of the pulse.  However, the spectrum of the radio pulse from a particle cascade depends on the angle from which it is viewed~\citep{alvarez-muniz2006}.  For this experiment, we chose a bandpass optimized for a cascade viewed from a point precisely on the Cherenkov cone, to minimize the threshold neutrino energy.  However, due to the small fractional bandwidth (300~MHz bandwidth divided by a center frequency of 1.35~GHz), applying this bandpass caused only a 1--2\% improvement in sensitivity.  Future experiments, with larger fractional bandwidths, will have larger potential gains from optimizing their bandpass.

\begin{figure}
 \centering
 \includegraphics[width=0.79\linewidth]{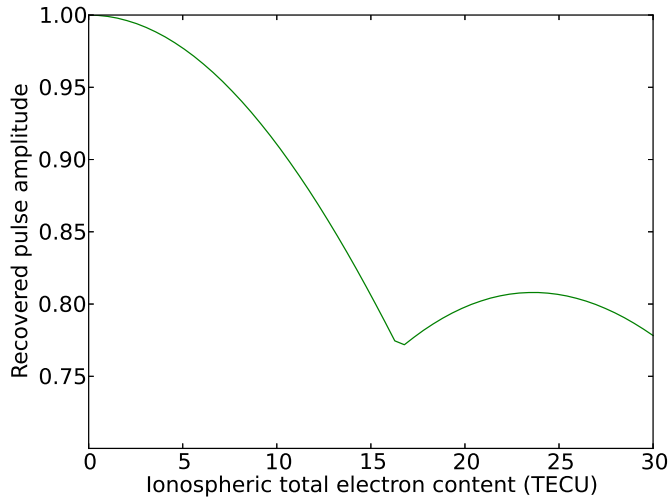}
 \caption{Effect of ionospheric dispersion on the fractional recovered amplitude of a pulse, assumed to have a flat spectrum across the frequency range \mbox{1.2--1.5~GHz}, after all other optimizations are applied.  The amount of dispersion is determined by the column density of electrons in the ionosphere, measured in Total Electron Content Units \mbox{(1~TECU $= 10^{16}$~electrons~m$^{-2}$)}.  Typical values in our experiment were \mbox{10--20~TECU}, and were corrected for by our dedispersion filter.}
 \label{fig:sigloss_disp}
\end{figure}

Variation in the phase of the transfer function corresponds to a delay in the expected pulse arrival time: linear variation causes a frequency-independent delay, which is unimportant; but quadratic and higher terms correspond to frequency-dependent dispersion, which affects the pulse amplitude.  The radio pulse from a particle cascade is not initially dispersed, but becomes so as it passes through the Earth's ionosphere.  For our experiment, the loss of sensitivity from this effect would be \mbox{$\sim 20\%$} (see \figref{fig:sigloss_disp}), were it not corrected by our dedispersion filter.

\begin{figure}
 \centering
 \includegraphics[width=0.79\linewidth]{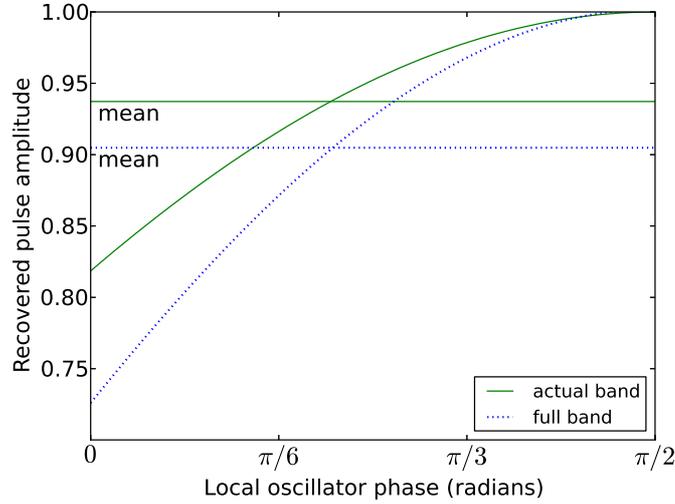}
 \caption{Effect of frequency downconversion on the fractional recovered amplitude of the radio pulse from a particle cascade, as a function of the phase of the local oscillator signal relative to the pulse.  When this phase is zero, the pulse shape is bipolar, minimising its peak amplitude.  The solid line is for our experiment, with an IF band of 50--350~MHz out of a possible 512~MHz; the dotted line is for an experiment which uses its entire IF band.}
 \label{fig:sigloss_phase}
\end{figure}

The constant term in the phase of the transfer function also relates to the shape of the expected pulse.  The radio pulse from a particle cascade is known to have a bipolar shape from both simulation~\citep{alvarez-muniz2010} and experiment~\citep{miocinovic2006}, indicating that the phase of the pulse is \mbox{$\pm \pi/2$}.  However, in our experiment we converted the signal from radio frequency (RF) to intermediate frequency (IF) by mixing it with a local oscillator signal: this adds the unknown phase of the local oscillator to the phase of the pulse, effectively randomising it, so it is not possible to design a filter to match it~\citep{bray2012}.  The mean loss of sensitivity from this effect is $\sim 6\%$ (see \figref{fig:sigloss_phase}); we corrected for this by calculating the signal envelope, which restores the full amplitude of the pulse, but this also increases the amplitude of the noise by $\sim 2\%$ (see \figref{fig:hist_byfilt}), so there is still a small net loss in the signal-to-noise ratio.

\begin{figure}
 \centering
 \includegraphics[width=0.79\linewidth]{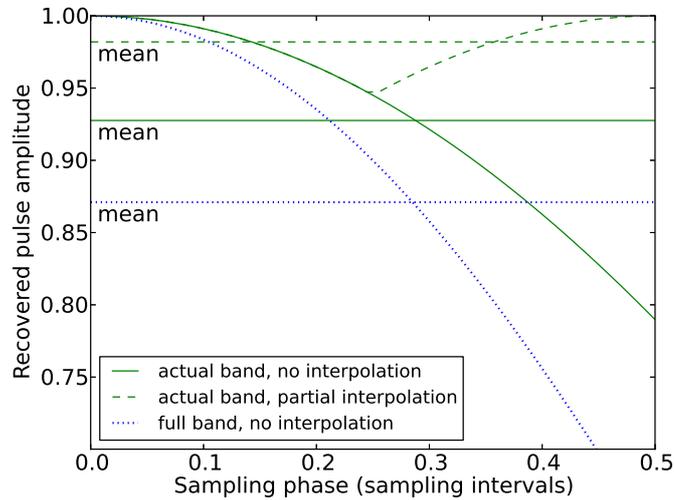}
 \caption{Effect of a finite sampling rate on the fractional recovered amplitude of a pulse, as a function of the offset between the sample times and the pulse peak.  The solid line is for the IF band of our experiment, with the dashed line incorporating the effects of the partial interpolation we performed in real time; the dotted line is for an experiment which uses its entire IF band.}
 \label{fig:sigloss_interp}
\end{figure}

The analog signal in a radio telescope is sampled only at a limited number of points, which will typically not record the peak amplitude of a pulse~\citep{james2010}.  However, the values of the signal between the sampled points, including the peak of the pulse, can be reconstructed through interpolation as a consequence of the Nyquist-Shannon sampling theorem.  We performed limited (two-fold) interpolation in real time, decreasing the mean loss of sensitivity from $\sim 7\%$ to $\sim 2\%$ (see \figref{fig:sigloss_interp}), with finer interpolation in later processing.

\section{Results from the Parkes experiment}
\label{sec:parkes_results}

\begin{figure}
 \centering
 \includegraphics[width=0.82\linewidth]{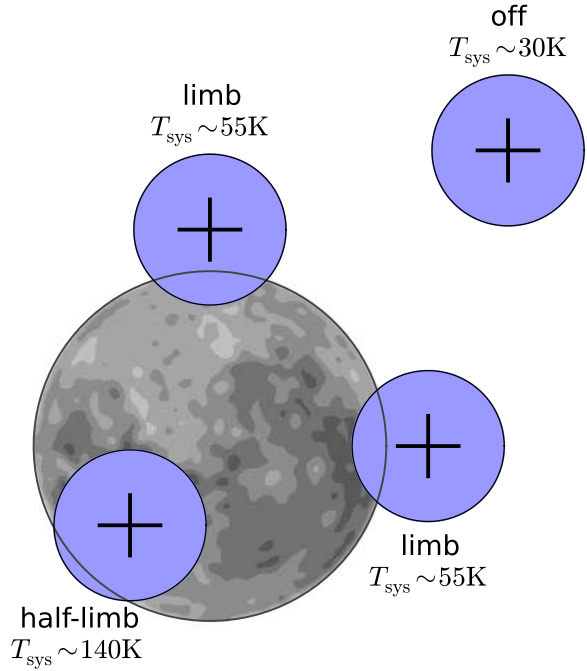}
 \caption{Example pointing of beams of the Parkes radio telescope relative to the Moon in our observations.  The limb beams receive less thermal radiation from the Moon than the half-limb beam, resulting in a lower system temperature $T_{\rm sys}$ and greater sensitivity to radio pulses.  The off-Moon beam is more sensitive still, but cannot detect neutrinos interacting in the lunar regolith: it is used only for anticoincidence RFI rejection.  Crosses indicate the alignment of the orthogonal linear polarizations of each beam; for each of the limb beams, one polarization is aligned radially to the Moon and the other tangentially.}
 \label{fig:pointing}
\end{figure}

The LUNASKA experiment with the Parkes radio telescope used the 21~cm multibeam receiver, with multiple beams directed on the limb of the Moon as shown in \figref{fig:pointing}.  Apart from increasing the neutrino aperture of the experiment compared to a single beam, this also provided a means to discriminate against RFI: a pulse from a lunar particle cascade should appear only in a single beam of the receiver, while RFI, entering through the antenna sidelobes, appears in all beams simultaneously.  To exploit this, we implemented an anticoincidence criterion in our backend hardware: an event, consisting of a high-significance peak in the measured signal voltage, triggered the storage of buffered data on all beams, but only if it exceeded the trigger threshold on only a single beam (within a 200~ns window).  The stored data were then fully optimized as described in \secref{sec:parkes_signal}, and subjected to a further series of cuts to exclude remaining RFI.

The first cut, which removed the majority of the remaining RFI, rejected events with either peaks in multiple beams or repeated peaks in a single beam, with a window extending out to several microseconds (the length of a buffer of stored data).  The RFI events removed by this cut were clustered on timescales of several seconds; assuming that RFI in the remaining events would also be clustered, we applied a second cut to reject all pairs of high-significance ($> 8\sigma$) events within 10~s of each other.  Finally, we applied a third, tighter anticoincidence cut, with a narrower time window (40~ns) allowing us to set a lower threshold for exclusion.

\begin{figure}
 \centering
 \includegraphics[width=0.96\linewidth]{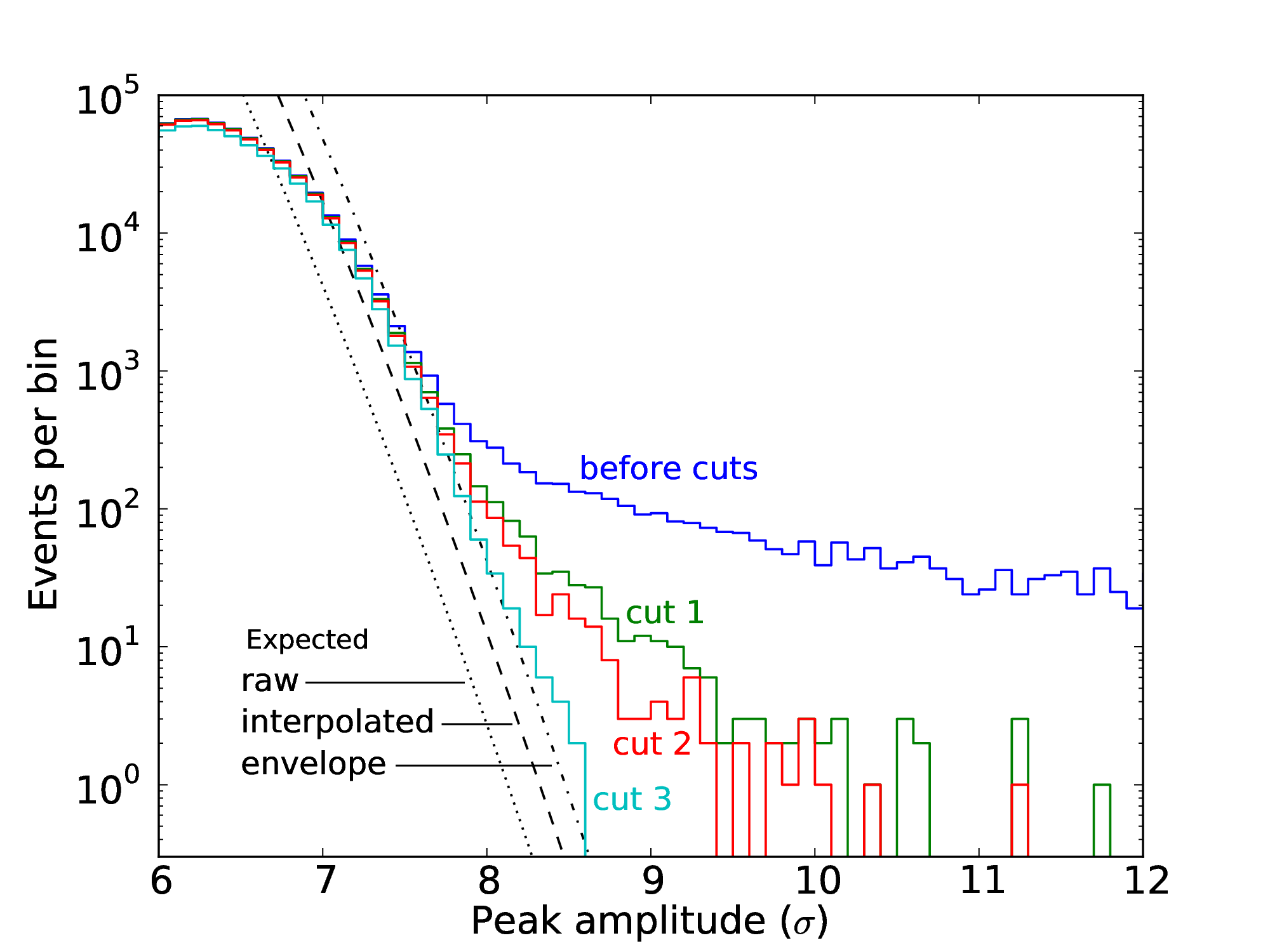}
 \caption{Histogram of peak amplitude (relative to the noise) in events recorded during the experiment with the Parkes radio telescope, before and after cuts to reject RFI (described in text), with a bin width of $0.1\sigma$.  Also shown are the expected numbers of peaks: from raw, sampled Gaussian noise; the same after interpolation; and after forming the signal envelope (which represents the processing actually performed on the data).  The deficit of recorded peaks at small amplitudes is due to incomplete triggering; the excess at large amplitudes is due to RFI, and is effectively removed by the cuts.}
 \label{fig:hist_byfilt}
\end{figure}

\begin{figure}
 \centering
 \includegraphics[width=\linewidth]{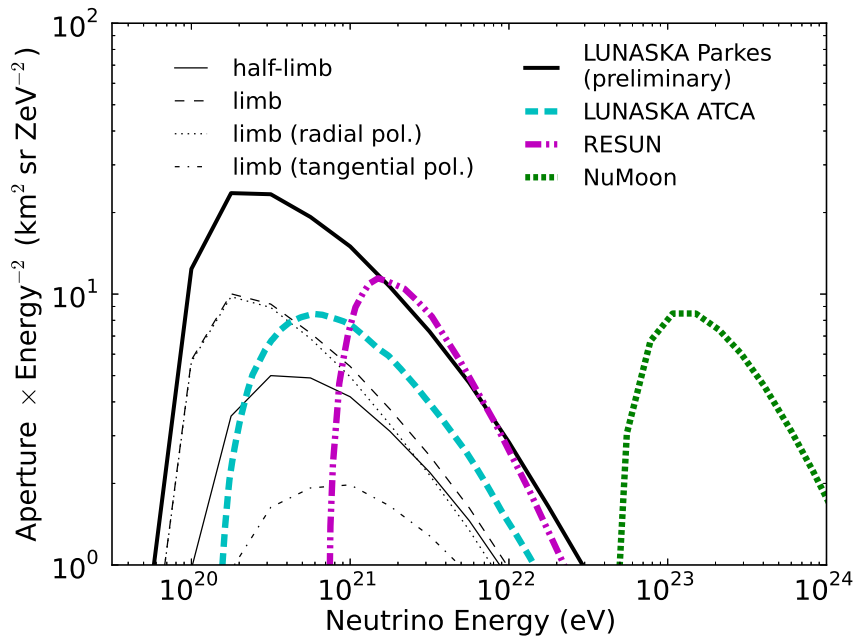}
 \caption{Neutrino apertures (thin lines) for the LUNASKA experiment with the Parkes radio telescope, for the individual beams shown in \figref{fig:pointing}.  The limb beam has a larger aperture than the half-limb beam partly due to its lower system temperature, and partly because it has one linear polarization aligned radially to the Moon, parallel to the expected polarization of the radio pulse from an interacting neutrino; the tangential polarization has a much lower aperture.  The combined aperture of all beams shown in \figref{fig:pointing} (thick solid line) is compared with the apertures of other recent lunar neutrino experiments: the previous LUNASKA experiment with the ATCA (for a limb pointing)~\citep{james2010}, RESUN~\citep{jaeger2010} and NuMoon~\citep{buitink2010}.  In this pointing configuration, the LUNASKA Parkes experiment has a greater instantaneous sensitivity to neutrinos than previous experiments in its energy range; the NuMoon experiment was sensitive to neutrinos at higher energies because it operated at a lower radio frequency.  Note that these apertures are based on different models: all LUNASKA curves are based on the simulations of \citet{james2009b}, the curve for RESUN is based on the analytic model of \citet{gayley2009}, and the NuMoon curve is based on the simulations of \citet{scholten2006}.  Data reduction for the joint Parkes-ATCA experiment is not yet complete, and the aperture for this experiment is not shown.}
 \label{fig:iso_ap}
\end{figure}

Histograms of peak amplitudes after each of these three cuts are shown in \figref{fig:hist_byfilt}.  Together, they rejected effectively all of the RFI, with no significant excess events remaining over those expected from random noise.  However, they had a \mbox{$\sim 7\%$} false rejection rate from the random noise, and an unknown false rejection rate for pulses of lunar origin with sufficient intensity to be detected through sidelobes of multiple beams.  Neglecting these effects, \figref{fig:iso_ap} shows a preliminary estimate of the neutrino aperture of our experiment.  A full analysis of our experiment's sensitivity, and the limit it places on the neutrino flux, is in preparation.




\section{Joint Parkes-ATCA experiment}
\label{sec:parkes-atca}

Our experiment using the Parkes and ATCA telescopes in combination aimed to establish the capability to detect a radio pulse with both telescopes simultaneously.  This would greatly improve the strength of RFI rejection: if a lone high-amplitude pulse were seen with the Parkes radio telescope alone, it would be unclear whether it originated from a lunar particle cascade or from local RFI that our cuts had failed to exclude; but the detection of a pulse with both of these telescopes (with a separation of 300~km), with the relative times of arrival indicating a lunar origin, would conclusively exclude the possibility of local RFI.

To exploit the full sensitivity of the ATCA, it is necessary to coherently combine the signals from its individual antennas, forming a tied-array beam.  To cover a significant fraction of the Moon, multiple such beams are required, which is impractical to achieve in real time.  Instead, we buffered the signal on each antenna individually, and stored the buffered data when triggered by the detection of a pulse by the Parkes radio telescope.  Tied-array beams can then be formed from these data in later processing to discover if the pulse was also detected by the ATCA.

\begin{figure}
 \centering
 \includegraphics[width=\linewidth]{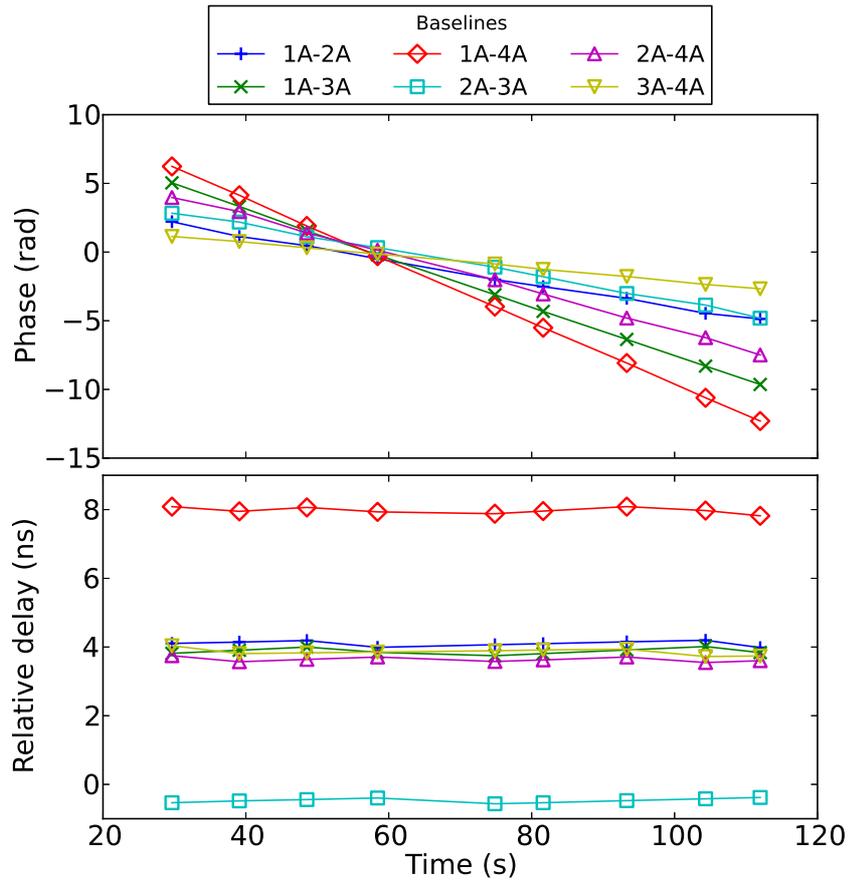}
 \caption{Phase and delay calibration for polarization A of all baselines between antennas 1--4 of the ATCA.  Phases show a linear drift, with the rate depending on the baseline length, as the calibration source moves across the sky.  Delays are roughly constant, due to the telescope's automatic delay correction.}
 \label{fig:atcacal}
\end{figure}

Two forms of calibration required for this experiment, made more difficult by the requirement that they be performed with the limited buffered data stored by the above procedure, are the delay calibration between the Parkes and ATCA telescopes, and between the separate antennas of the ATCA.  For the first, we used the BeiDou\mbox{-}1C satellite as a calibrator source, achieving a timing precision of $\sim 50$~ns, which is more than sufficient to determine whether a pulse originated from the Moon~\citep{bray2011c}.  For the second, we used the astronomical source 3C273, and calibrated the phases and delays as shown in \figref{fig:atcacal}.

The ATCA applies its own approximate delay correction, to the nearest sampling interval (0.25~ns), which is responsible for the pattern seen in \figref{fig:atcacal}.  The next stage in the data reduction for this experiment is to remove this approximate correction and replace it with a more precise correction, allowing the coherent combination of the ATCA antennas to form a tied-array beam on the Moon.

\section{Future prospects}
\label{sec:future}

As the sensitivity of available radio instruments determines the sensitivity of lunar neutrino experiments, it is worthwhile to look at current developments in this field which may be useful in this application.  One such development is the phased array feed (PAF), such as the design developed for the Australian SKA Pathfinder (ASKAP), which samples the electric field at multiple points in a telescope's focal plane and forms beams through later digital processing.  If a PAF were mounted on the Parkes radio telescope, as has been proposed for other astronomical applications, beams could be formed around the entire limb of the Moon rather than the limited coverage shown in \figref{fig:pointing}, allowing a \mbox{$\sim 10\times$} increase in neutrino aperture.  The PAFs developed for ASKAP also have a frequency range of 0.7--1.8~GHz, which is a substantial improvement over the 1.2--1.5~GHz of the current Parkes 21~cm multibeam receiver; if fully processed, this increased bandwidth would allow a \mbox{$\sim 2\times$} decrease in the threshold neutrino energy.


Looking further into the future, the site decision for the SKA, which will be a substantial improvement over all current instruments, was made in May~2012: its high-frequency component will be located in South Africa and its low-frequency component at the Murchison Radio-astronomy Observatory (MRO) in Australia, both to start construction in 2016.  The details of its signal path and its suitability for the nanosecond-scale pulse detection required in this application are yet to be determined; but its low-frequency precursor at the MRO, the Murchison Widefield Array (MWA), is likely to influence its design.

\section{Conclusion}

The LUNASKA project has conducted a series of experiments with the Parkes and ATCA radio telescopes to search for UHE neutrinos using the Moon as a target volume.  We have developed signal processing techniques to maximize the sensitivity of such experiments, which will be required in order to fully realize the potential of future radio telescopes to detect neutrinos at the very highest energies.

\section*{Acknowledgements}

The Australia Telescope Compact Array and the Parkes radio telescope are both part of the Australia Telescope which is funded by the Commonwealth of Australia for operation as a National Facility managed by CSIRO.  This research was supported by the Australian Research Council's Discovery Project funding scheme (project number DP0881006).

\bibliographystyle{aipproc}
\bibliography{all}

\end{document}

%% file: journals.tex
\def\arxiv{ArXiv e-prints}    
\def\aspacer{Adv.~Space R.}   
\def\aap{A\&A}                
\def\aapr{A\&A~Rev.}          
\def\aaps{A\&AS}              
\def\aj{AJ}                   
\def\ajph{Australian J.~Phys.}
\def\alet{Astro.~Lett.}       
\def\anchem{Analytical~Chem.} 
\def\ao{Applied Optics}       
\def\apj{ApJ}                 
\def\apjl{ApJ}                
\def\apjs{ApJS}               
\def\app{Astropart.~Phys.}    
\def\apss{Ap\&SS}             
\def\araa{ARA\&A}             
\def\arep{Astron.~Rep}        
\def\aspconf{Astron.~Soc.~Pac.~Conf.} 
\def\asr{Av.~Space Res.}      
\def\azh{AZh}                 
\def\baas{BAAS}               
\def\bell{Bell~Systems~Tech.~J.} 
\def\cpc{Comput.~Phys.~Commun.} 
\def\epsl{Earth and Plan.~Sci.~Lett.} 
\def\expa{Exp.~Astron.}       
\def\gca{Geochim.~Cosmochim.~Acta} 
\def\grl{Geophys.~Res.~Lett.} 
\def\iaucirc{IAU Circ.}       
\def\ibvs{IBVS}               
\def\icarus{Icarus}           
\def\ieeetit{IEEE Trans.~Inf.~Theor.} 
\def\ijmpd{Internat.~J.~Mod.~Phys.~D} 
\def\invp{Inverse Prob.}      
\def\jastp{J.~Atmos.~Solar-Terr.~Phys.} 
\def\jcap{J.~Cosm.~Astropart.~Phys.} 
\def\jcomph{J.~Comput.~Phys.} 
\def\jcp{J.~Chem.~Phys.}      
\def\jewa{J.~Electromagn.~Wav.~Appl} 
\def\jgeod{J.~Geodesy}        
\def\jgr{J.~Geophys.~R.}      
\def\jhep{JHEP}               
\def\jrasc{JRASC}             
\def\met{Meteoritics}         
\def\mmras{MmRAS}             
\def\mnras{MNRAS}             
\def\moonp{Moon and Plan.}    
\def\mpla{Mod.~Phys.~Lett.~A} 
\def\mps{Meteoritics and Planetary Science} 
\def\nast{New Astron.}        
\def\nat{Nature}              
\def\nima{Nucl.~Instrum.~Meth.~A} 
\def\njp{New J.~Phys.}        
\def\nspu{Phys.~Uspekhi}      
\def\pasa{PASA}               
\def\pasj{PASJ}               
\def\pasp{PASP}               
\def\phr{Phys.~Rev.}          
\def\pla{Phys.~Lett.~A}       
\def\plb{Phys.~Lett.~B}       
\def\pop{Phys.~Plasmas}       
\def\pra{Phys.~Rev.~A}       
\def\prb{Phys.~Rev.~B}        
\def\prc{Phys.~Rev.~C}        
\def\prd{Phys.~Rev.~D}        
\def\prl{Phys.~Rev.~Lett.}    
\def\pst{Phys.~Scr.~T}        
\def\phrep{Phys.~Rep.}        
\def\phss{Phys.~Stat.~Sol.}        %
\def\procspie{Proc.~SPIE}     
\def\planss{Planet.~Space Sci.}  
\def\qjras{QJRAS}             
\def\radsci{Radio Sci.}       
\def\rpph{Rep.~Prog.~Phys.}   
\def\rqe{Rad.~\&~Quan.~Elec.} 
\def\rgsp{Rev.~Geophys.~Space Phys.~} 
\def\rsla{Royal Soc.~Trans.~A} 
\def\sal{Sov.~Astron.~Lett.}
\def\spjetp{Sov.~Phys.~JETP}  
\def\spjetpl{Sov.~Phys.~JETP~Lett.} 
\def\spu{Sov.~Phys.~Uspkehi}  
\def\sci{Science}             
\def\solph{Sol.~Phys.}        
\def\ssr{Space Sci.~Rev.}     
\def\wars{Workshop on App.~of Radio Sci.} 
\def\zap{Z.~Astrophys.}       